\RequirePackage{snapshot}
\documentclass[aps,prx,twocolumn,amsfonts,showpacs,longbibliography]{revtex4-2}
\usepackage{verbatim,epsfig,amsmath,amssymb,bm,epsf,graphicx,psfrag,bbold,amsthm,amsfonts}
\usepackage[toc,page,titletoc]{appendix}
\usepackage[bottom]{footmisc}
\usepackage{hyperref}
\hypersetup{
    colorlinks=true,
    linkcolor=black,
    citecolor=blue,
    filecolor=black,
    urlcolor=blue,
}

\usepackage[capitalize]{cleveref} %should be placed only after hyperred
\usepackage{enumitem}
\usepackage{soul}
\usepackage{grffile}
\usepackage{framed}
\usepackage{mathrsfs}
\usepackage{esint}
\setlist{nosep}
\usepackage{placeins}
\usepackage[all]{xy}
\usepackage{color}
\usepackage[utf8]{inputenc}
\usepackage{float}
\usepackage{natbib}
\usepackage{tikz-cd}
\usepackage{verbatim}
\usepackage{leftidx}
\usepackage[normalem]{ulem}

\usepackage{notes2bib}

\newcommand{\abs}[1]{\left|#1\right|}

\newcommand{\cZ}{\mathcal{Z}}

\newcommand\bR {{\mathbb R}}

\newcommand\beq {\begin{equation}}
	\newcommand\eeq {\end{equation}}
\newcommand\beqa {\begin{equatiobn}\begin{array}}
		\newcommand\eeqa {\end{array}\end{equation}}
\newcommand\bal {\begin{align}}
	\newcommand\eal {\end{align}}
\newcommand{\bea}{\begin{eqnarray}}
	\newcommand{\eea}{\end{eqnarray}}

\theoremstyle{plain}

\theoremstyle{definition}

\theoremstyle{remark}

\usetikzlibrary{arrows,decorations.pathreplacing}
\usetikzlibrary{decorations.pathmorphing}

\tikzset{snake it/.style={decorate, decoration=snake}}\usepackage[T1]{fontenc}

\usetikzlibrary{decorations.markings}

\begin{document}
	
	\title{Machian fractons, Hamiltonian attractors and non-equilibrium steady states}

	\author{Abhishodh Prakash}
	\email{abhishodh.prakash@physics.ox.ac.uk (he/him/his)}
	\affiliation{Rudolf Peierls Centre for Theoretical Physics, University of Oxford, Oxford OX1 3PU, United Kingdom}
 \author{Ylias Sadki}
 	\email{ylias.sadki@physics.ox.ac.uk}
	\affiliation{Rudolf Peierls Centre for Theoretical Physics, University of Oxford, Oxford OX1 3PU, United Kingdom}
	\author{S. L. Sondhi}
	\email{shivaji.sondhi@physics.ox.ac.uk}
	\affiliation{Rudolf Peierls Centre for Theoretical Physics, University of Oxford, Oxford OX1 3PU, United Kingdom}

	\begin{abstract}
            We study the $N$ fracton problem in classical mechanics, with fractons defined as point particles that conserve multipole moments up to a given order. We find that the nonlinear Machian dynamics of the fractons is characterized by late-time attractors in position-velocity space for all $N$, despite the absence of attractors in phase space dictated by Liouville's theorem. These attractors violate ergodicity and lead to non-equilibrium steady states, which always break translational symmetry, even in spatial dimensions where the Hohenberg-Mermin-Wagner-Coleman theorem for equilibrium systems forbids such breaking. We provide a conceptual understanding of our results using an adiabatic approximation for the late-time trajectories and an analogy with the idea of `order-by-disorder' borrowed from equilibrium statistical mechanics. Altogether, these fracton systems host a new paradigm for Hamiltonian dynamics and non-equilibrium many-body physics.
        \end{abstract}
	
	\maketitle
 \tableofcontents
   
       \section{Introduction} 
       The notion of thermal equilibrium and the technology of statistical mechanics are central to our understanding of macroscopic systems. The idea of ergodicity bridges the intellectual gap between the unceasing microscopic evolution of any system and the success of time-independent statistical averages. If the system dynamics is ergodic, the properties of the late-time states reached by starting from generic initial conditions should agree~\footnote{We will refer to this as equilibration without the qualifying ``thermal''.} and be described by statistical mechanics. For classical systems---and this is a paper about those---the canonical picture of ergodicity is that while the precise details of a particular trajectory depend sensitively on initial conditions, a typical trajectory densely covers all phase space available to it, consistent with conservation laws, by repeatedly revisiting the vicinity of any allowed phase space point under dynamics~\footnote{We are being purists here. In practice, a system may be ``ergodic enough for government work'' and given experimental times, this is not a distinction that can be tested directly. We are also unaware of a usable definition of a system that is ``ergodic enough''.}.

    The question of deciding whether a given Hamiltonian gives rise to ergodic dynamics or not has a long and distinguished history. For macroscopic systems one tends to assume that ergodicity is the norm unless the system is explicitly integrable and that integrable systems are isolated points in Hamiltonian space. In this paper, we describe a family of Hamiltonian systems whose native physics violates this expectation and leads to a breakdown of equilibration and statistical mechanics. These are systems of fractons which have been the subject of a large volume of work in recent years in the quantum mechanical setting~\cite{Chamon_Fracton_PhysRevLett.94.040402,Haah_FractonPhysRevA.83.042330,VijayHaahFu_Fractons_PhysRevB.92.235136,NandkishoreHermeleFractonsannurev-conmatphys-031218-013604,PretkoChenYou_2020fracton,GromovRadzihovsky2022fractonReview} but whose classical mechanics has only recently been introduced and studied by two of us and Goriely~\cite{AP2023NRFractons} for small numbers of particles. More precisely, we consider ``ungauged'' fractons, i.e. particles whose dynamics conserve a consistent set of charge multipoles. Symmetry and locality dictate that such particles obey ``Machian'' dynamics, where their inertial response to forces depends entirely on their proximity to other particles, unlike Newtonian dynamics, where it depends entirely on a property (the mass) of the particle alone.

      \begin{figure}[!ht]
            \centering
            \includegraphics[width=8.6cm]{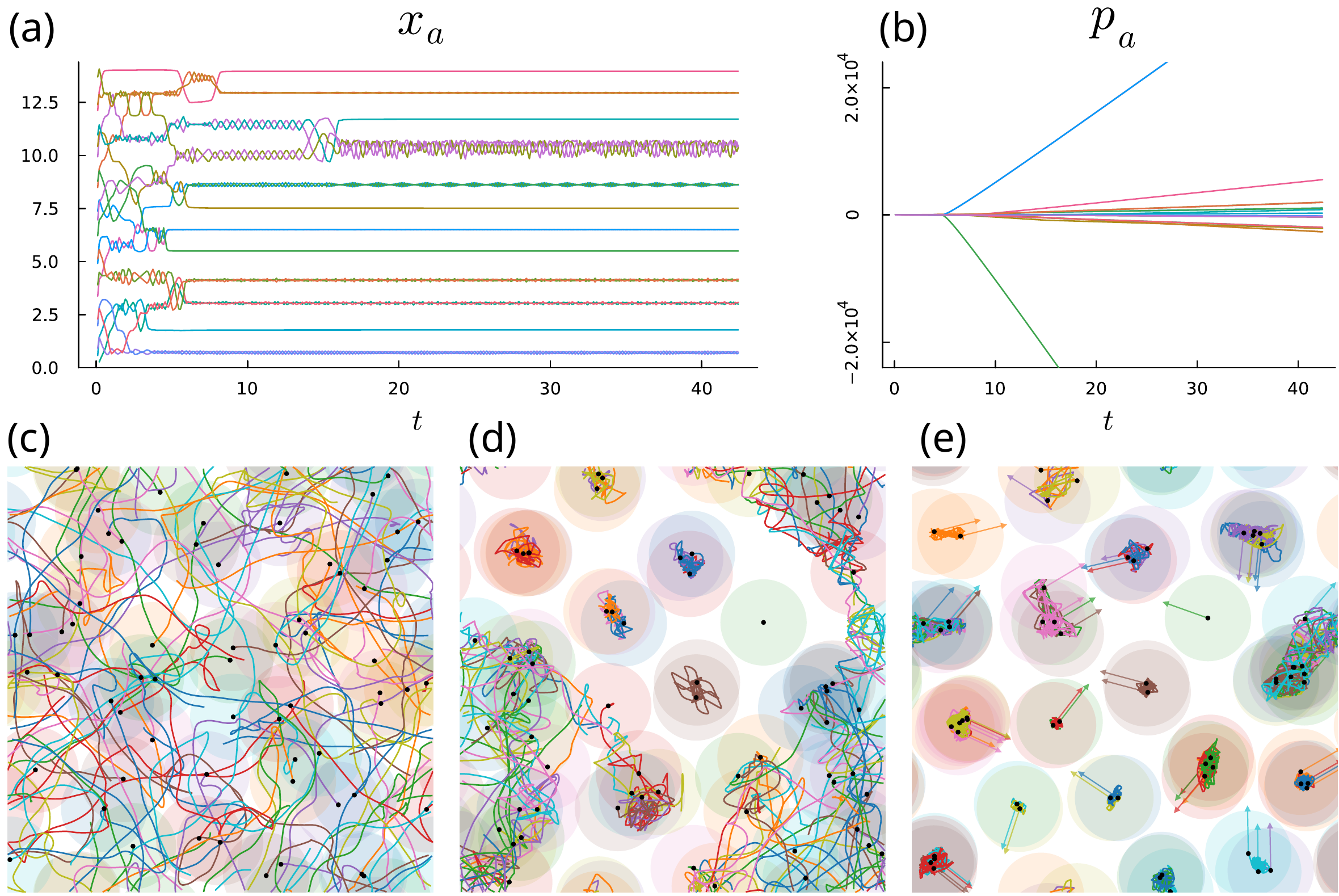}
            \caption{
            (a)-(b): Position ($x_a$) and momentum ($p_a$) fractonic trajectories of a 1d system of 20 particles starting with uniform density $\rho = 1.4$.
            (c)-(e): Position trajectories of a 2d system of 64 particles starting from uniform density $\rho = 2.56$. Shaded circles with radius $0.5$ are drawn around each particle for visual clarity. The directions of the momenta are included as arrows in (e). All trajectories are generated by the Hamiltonian in \cref{eq:H_dipole} with $g=0.3$. Both the 1d and 2d systems exhibit emergent crystallization starting from random initial conditions.
            }
            \label{fig:main}
        \end{figure}
    Machian dynamics, in turn, dictates a remarkable set of properties for systems of $N$ particles. First, their motion converges at late times to attractors. Naively this should be impossible in a Hamiltonian system obeying Liouville's theorem, but the attractors are in position-velocity space instead of in phase space, and the relationship between velocities and momenta is very different in Machian and Newtonian dynamics. Second, there are many attractors for large $N$ and so the dynamics does {\it not} lead to late-time states whose properties are governed solely by global conserved quantities. Instead, we see the emergence of further conserved quantities at asymptotically late times. Third, late-time states {\it always} break translation symmetry even in low dimensions, where the naive invocation of the Hohenberg-Mermin-Wagner-Coleman~\cite{HohenbergPhysRev.158.383,MerminWagner_PhysRevLett.17.1133,Colemancmp/1103859034} theorem would forbid breaking of this continuous symmetry. Perhaps most striking (see \cref{fig:main}) is the frequent evolution of high-density fractons systems into states with crystalline order!

To characterize the non-linear dynamics of our system, we perform a stability analysis on the solution space. 
We develop an asymptotically self-consistent separation into fast and slow variables that demonstrates the existence of attractors.
Further, we show that partition functions for our systems are generically divergent due to the non-compactness of energy hypersurfaces, consistent with the breakdown of equilibration observed in dynamics.
Despite this divergence, statistical mechanical reasoning of the kind used in ``order by disorder'' (OBD)~\cite{MoessnerChalkerOBDPhysRevLett.80.2929,ChalkerOBD2011} discussions in ergodic systems can be adapted to gain insight into the temporal evolution of our non-ergodic system.
Essentially, the divergences stem from zero modes in phase space whose numbers depend on particle configurations in real space, and the observed dynamics tends to maximize this number. 

In this paper, we provide the technical content of the above assertions. Before we do that, we remark that much recent work on quantum systems has focused on the breakdown of quantum ergodicity---most closely in lattice fracton systems at low density in the phenomenon termed ``shattering'' of Hilbert space and most famously in the phenomenon of many body localization~\cite{NandkishoreHuse_MBLReview} in disordered systems. Although our classical systems are very far from these strongly quantum systems with very small local Hilbert spaces, it is nonetheless notable that we find analogs of shattering in our multiple attractor dynamics and of localization-protected quantum order~\cite{HuseNandkishorePalSondhi_LocalizationProtectedOrderPhysRevB.88.014206} in the breakdown of translation invariance.

\section{Symmetries and Hamiltonians}
\begin{figure}
    \centering
    \includegraphics[width = 8.6cm]{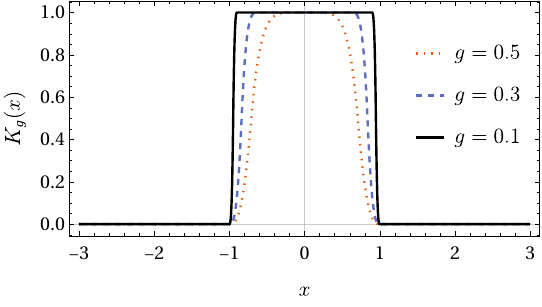}
    \caption{The locality function $K_g(x)$ defined in \cref{eq:K} for various representative values of $g$.}
    \label{fig:K}
\end{figure}
We consider $N$ identical non-relativistic point particles in $d$ spatial dimensions. The state of the system is specified by $Nd$ coordinates, $\{x^\mu_j, p^\mu_j \}$ where the Greek superscript indices $\mu=1,\ldots,d$ denote the component and the Latin subscript indices $j = 1,\ldots,N$ denote the particle number. We will be interested in two classes of symmetries. The first is spatial translation, which acts on position coordinates as $x^\mu_j \mapsto x^\mu_j + \alpha^\mu$ and leads to the conservation of the total momentum, $P^\mu = \sum_j p^\mu_j$. The second is the conservation of the total multipole moment $Q^\mu_{\ell} \equiv \sum_j (x^\mu_j)^\ell$. We will focus on $\ell =1$ for now, when $Q^\mu_1 \equiv D^\mu$ denotes the dipole moment. Dual to translations, $D^\mu$ generates rigid shifts of the momentum coordinates~\cite{AP2023NRFractons} as $p^\mu_j \mapsto p^\mu_j + \beta^\mu$. A physically sensible and local Hamiltonian compatible with both symmetries takes the form~\cite{AP2023NRFractons}

\begin{equation}
            H = \sum_{a<b=1}^N \frac{\left(\vec{p}_a - \vec{p}_b \right)^2}{2} K(\abs{\vec{x}_a - \vec{x}_b}) + \ldots \label{eq:H_dipole}
\end{equation} 
where $K(x)$ is a positive `mobility function' that imposes locality. The ellipses in \cref{eq:H_dipole} indicate other local symmetric terms, including conventional interactions, which we drop in this work for simplicity as their effects do not qualitatively modify those we report~.
       
In this work, we require $K(x)$ to have a strictly compact support restricted to $|x| \le l_M$ where 
 the Machian length $l_M$ is a microscopic length scale that characterizes dynamics~\footnote{We discuss at the end of the paper what happens when we loosen this restriction.}. It is useful to pick families of functions that contain as a limit the indicator function on this interval. In this paper, we use the following family with Machian length $l_M$ set to 1:
\begin{equation}
     K_g(x) = \begin{cases}
      1 & \mbox{ $ x^2 \leq 1-2g $}\\
      1 - \frac{1}{2g^3}(x^2 -1 +2g)^3 & \mbox{ $ 1-2g < x^2 \leq 1-g $}\\
      - \frac{1}{2g^3}(x^2 -1)^3 & \mbox{ $ 1-g < x^2 < 1 $}\\
      0 & \mbox{ $ x^2 \geq 1 $}.
      \end{cases} \label{eq:K}
\end{equation}
     \Cref{eq:K}  is continuous and differentiable and takes on the desired limiting form
\begin{equation}
       \lim_{g \rightarrow 0} K_g (x) = \Theta(x+1) - \Theta(x-1). \label{eq:K_box}
\end{equation}
where $\Theta(x)$ is the Heaviside function.

\section{$N$ particle dynamics}
It is clear from the form of \cref{eq:H_dipole} that the dynamics is Machian. $H$ vanishes for isolated, immobile, particles and mobility is restored only by the proximity of others within a Machian length $l_M$. The few-body dynamics of \cref{eq:H_dipole} for $N \le 6$ was studied in \cite{AP2023NRFractons} where it was shown that particles initialized in sufficient proximity generically separate into multiple clusters. The centers of mass of the clusters become immobile and behave as asymptotic conserved quantities, while particles within a cluster with more than one particle exhibit oscillations. The position- velocity space exhibits attractors in the form of stable fixed points and limit cycles, while there are no attractors in phase space, in conformity with Liouville's theorem.
        
We now turn to the dynamics generated by \cref{eq:H_dipole} for the finite-density problem of interest for macroscopic systems, i.e. the limit $N\rightarrow \infty$ and volume $V = L^d \rightarrow \infty$ keeping $\rho = N/V$ fixed. Although we first focus on one dimension for simplicity, we find analogous phenomena in higher dimensions. 

We focus on random initial conditions at fixed energy. Particles are distributed uniformly in space, with momenta chosen by a random walk in momentum space, terminated when the desired energy is obtained.
Subsequently, we numerically solve the Hamilton equations. For example, the plots in \cref{fig:main} were generated this way with $K(x)$ in \cref{eq:K} with $g=0.3$. Our principal findings are as follows:

\smallskip

        \begin{enumerate}[wide, labelindent=0pt]
        \itemsep0.5em
            \item For low densities $\rho < l^{-1}_M$, generic random initial conditions lead to locations of particles distributed as a Poisson process with mean nearest-neighbor separation $\sim \rho^{-1} > l_M$ and results, with high probability, in isolated particles lacking any neighbors within $l_M$. 
            All of the energy resides in relatively rare active groups,
            which splinter and form multiple steady-state clusters as discussed in \cite{AP2023NRFractons}.
            
            \item  For high densities $\rho> l^{-1}_M$ the mean nearest-neighbor separation is now less than $l_M$. Thus, most particles start off within a large active group that potentially spans the system, seemingly favoring restoration of ergodicity for generic initial conditions. Indeed, quantum lattice fractons 
            \cite{SkinnerPozderac_Thermalization_2023,MorningstayKhemaniHust_Thermalization_PhysRevB.101.214205} exhibit such a restoration of ergodicity. Surprisingly, this does {\it not} happen in our models. Instead, we continue to see ergodicity breaking and the formation of clusters with $\approx \rho$ number of particles each, but now spaced at regular intervals of distance $\approx l_M$. The distribution of particles among the clusters fluctuates with initial conditions. The trajectories of a high-density 1d system are shown in \cref{fig:main}(a),(b).            
            \item With big bang initial conditions at high density, the particles generically do not go on to occupy all the position space but remain localized within a finite number of clusters, each with a large number of members (see \cref{app:initial_condition_dependence}). 
            \item  These observations are generalized straightforwardly to higher dimensions, as shown in \cref{fig:main}(c)-(e). 
           
        \end{enumerate}

        \begin{figure}
            \centering
            \includegraphics[width=8.6cm]{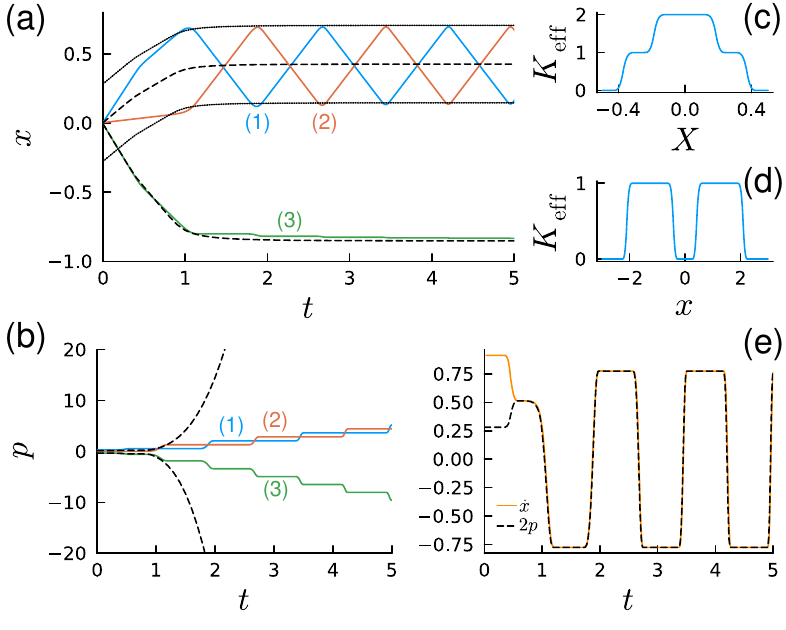}
            \caption{
            (a,b): Position and momentum trajectories of a 3 particle system (solid colored lines) against adiabatic solutions for the `slow' variables $X,P$ (dashed lines) obtained from solving \cref{eq:Adiabatic_XP}.  The $X$ trajectory accurately reproduces the motion of the cluster centers, whereas the $P$ trajectory deviates from it as errors propagate due to its divergent nature. Dotted lines indicate $X \pm \epsilon$ where $\epsilon=0.28$ is fitted to match the amplitude of the oscillating fast variable $x$. 
            (c), (d): $K_{\mathrm{eff}}(x)$ defined in \cref{eq:Keff} plotted as a function of $X$ where it represents an effective mobility function and $x$ where it represents a confining potential.
            (e): $\dot{x}=(\dot{x}_1 - \dot{x}_2)/2$ and $2p = p_1 - p_2$  from the exact trajectory, predicted to match in the adiabatic solution \cref{eq:Adiabatic}.
            }
            \label{fig:3particles}
        \end{figure}

    \section{Broken Ergodicity and the Hohenberg-Mermin-Wagner-Coleman theorem} At any density the state of the fractons, with global conserved quantities fixed, converges to one of a large number of attractors, all of which spontaneously break translation symmetry. This occurs even in $d=1,2$ where the theorem of Hohenberg, Mermin, Wagner, and Coleman (HMWC)~\cite{HohenbergPhysRev.158.383,MerminWagner_PhysRevLett.17.1133,Colemancmp/1103859034} forbids the breaking of continuous symmetries in classical systems at nonzero energy densities~\footnote{Recent work~\cite{Kapustin_HMW_PhysRevB.106.245125,RahulNandkishore_HMW_PhysRevB.105.155107} has shown that the HMWC analysis needs to be modified when applied to multipole symmetry breaking. As far as we can tell, this does not alter the conditions for translation symmetry breaking.}. 
    Evidently the theorems are evaded by breaking ergodicity and hence the assumption of validity of statistical mechanics (or equivalent Euclidean quantum mechanics).
    Interestingly, a similar way around the theorems was discovered in~\cite{HuseNandkishorePalSondhi_LocalizationProtectedOrderPhysRevB.88.014206} that involved one of us. There it was shown that for quantum many-body systems with strong quenched disorder that break ergodicity and exhibit many-body localization (MBL), discrete symmetries~\cite{PotterVasseur_symmetryPhysRevB.94.224206,AP_S3MBL_PhysRevB.96.165136} can be broken in highly excited states with finite energy density even in $d=1$---a phenomenon termed localization-protected quantum order (LPQO). In our case, the mechanism leading to broken ergodicity is entirely different, so even a continuous symmetry can be broken.

We now turn to providing a conceptual understanding of the above results. First we  provide a conceptual understanding of the above results in terms of a self-consistent treatment at late times. Thereafter, we will turn to a statistical mechanical perspective.

\section{Machian Schismogenesis} We now analyze the cluster formation for the N particle problem, for which the three-particle system is a tractable microcosm. As seen in \cref{fig:3particles}, particles that start within a single active group generically splinter into two clusters. Once this sets in, there is a seeming barrier to further inter-cluster exchanges. We will now show that this is intimately related to the dynamics in the momentum space, shown in \cref{fig:3particles}(d). As clustering sets in, the momenta branch out and evolve in such a way that momentum differences between particles within a cluster are finite, whereas across the clusters they diverge with time. The latter generates the barrier observed in position-space dynamics. To see this, let us look at the equations of motion for the three-body Hamiltonian~\cref{eq:H_dipole},
    \begin{align}
        \dot{x}_a &= \sum_{b\neq a} (p_a - p_b) K(x_a - x_b),\\
        \dot{p}_a &= -\sum_{b\neq a} \frac{(p_a - p_b)^2}{2} K'(x_a - x_b). \label{eq:EOM_1d_dipole}
    \end{align}
    The important observation is that the dynamics in \cref{fig:3particles} occurs along two time scales, a slow one by the centers of clusters and a fast one within the two-particle cluster where particles oscillate with a small amplitude and high frequency. It is convenient to employ a corresponding decomposition of phase-space variables to reflect this
    \begin{align}
        x_1 &= X +x,~ x_2 = X-x,~ x_3 = -2X, \nonumber\\
        p_1 &= P +p,~ p_2 = P-p,~ P_3 = -2P. \label{eq:XP}
    \end{align}
    The upper and lower case variables are the slow and fast degrees of freedom, respectively. The phase-space variables are not all independent since the total momentum and positions are conserved, and we have fixed both to zero without loss of generality. When momentum branching sets in, we will show that assuming $P>>p$ and $|x| <<1$ produce a self-consistent solution where the equations of motion in \cref{eq:EOM_1d_dipole} are simplified to (see \cref{app:Adiabatic}):
    \begin{align}
        \dot{X} &\approx \frac{3P}{2}K_{\rm eff}(X,x), ~\dot{P} \approx-\frac{9P^2}{4} \frac{\partial K_{\rm eff}(X,x)}{\partial x},\label{eq:Adiabatic_XP}\\
        \dot{x} &\approx 2p,~\dot{p}  \approx -\frac{9P^2}{4} \frac{\partial K_{\rm eff}(X,x)}{\partial x}. \label{eq:Adiabatic}
    \end{align}
    $K_{\rm eff}(X,x)$ is defined as
    \begin{equation}
        K_{\rm eff}(X,x) \equiv K(3X+x) +K(3X-x).
        \label{eq:Keff}
    \end{equation}
    Now we analyze \cref{eq:Adiabatic} in the adiabatic approximation (see \cite{LandauVol11982mechanics} and \cref{app:Adiabatic}). First, we treat the slow variables $X,P$ as constants and solve the equations for $x,p$ whose motion corresponds to an effective Newtonian particle with mass $0.5$ in an external potential. 
    \begin{equation}
        H_{X,P}(x,p) \approx p^2 + \frac{9P^2}{4}  K_{\rm eff}(X,x). \label{eq:Heff_xp}
    \end{equation}
    Assuming that $3X = 1+ \xi$ for some $1>\xi>0$, the potential takes the form shown in \cref{fig:3particles}. This strongly confines the particle within $|x| < |\xi|$ where the particle oscillates rapidly with amplitude $\xi$. We now feed in the time-averaged fast solution into the equations for $X,P$ simply by replacing $x \rightarrow \xi$. This slow motion is generated by the Hamiltonian
    \begin{equation}
        H_\xi(X,P) \approx \frac{3P^2}{4} K_{\rm eff}(X,\xi). \label{eq:Heff_XP}
    \end{equation}
    \Cref{eq:Adiabatic_XP} was studied in~\cite{AP2023NRFractons} (see also \cref{app:Adiabatic}) and describes the dynamics of a pair of fractons, here to be understood as the cluster centers. At late times, the system in \cref{eq:Adiabatic_XP} reaches a steady state with $\dot{X} \rightarrow 0$, $P$ diverging and $X$ taking the smallest value so that $K_{\rm eff}(X) \rightarrow 0$. For the form shown in \cref{eq:Keff}, this corresponds to $|3X| = 1+ \xi$, which self-consistently supports the earlier assumption for fast motion. With time, we see that the solution with the adiabatic approximation is increasingly valid: (i) the cluster centers freeze out at positions $X$ and $-2X$ while the particles within a cluster rapidly oscillate with amplitude $\sim \xi$. The precise value of $|\xi|<1$ and the oscillation frequency depend on the initial conditions. In \cref{fig:3particles} we compare the adiabatic solution for $X,P$ (broken lines) with the actual dynamics (solid colored lines). We see that by fitting $\xi$ to the amplitude of the fast oscillation (bounds of dotted lines), we obtain excellent agreement with the motion of the centers of clusters for late times. Furthermore, the dynamics within a cluster satisfies the Newtonian relation $\dot{x}  = 2p$ as expected from \cref{eq:Adiabatic}. The solution for the momentum $P$ deviates substantially from the adiabatic solution, as the errors are compounded due to its divergent nature. However, from our perspective, the main quantitative physics is in position space, whereas the momentum-space behavior is important only qualitatively, which the adiabatic approximation nicely reproduces.  

    This calculation can also be distilled into a more intuitive understanding by tracking how energy is distributed. Once clustering sets in, the energy is carried mainly by the active pair $1-2$, $E = \frac{1}{2}(p_1 - p_2)^2 = 2 p^2$ and does not depend on the large values of $P$. However, when one of these particles, say $1$ approaches $3$, the energy cost of the two entering each other's range is $\delta E \sim \frac{1}{2}(3P+ p)^2$. Thus, $1$ senses a large energy barrier and is repelled, while $2$ reverses its motion to conserve the center of mass, and the story repeats. With time, as $P$ increases, so does this energy barrier to cluster restructuring, and the particles are confined to their clusters. Although all \emph{physical} attributes, such as positions and velocities, are comparable for all three particles,  irreconcilable momentum differences make cluster identities asymptotically immutable. We term this \emph{Machian schismogenesis}, after a similar social phenomenon~\cite{Bateson1935culture}. 

    The generalization to larger number of clusters and higher dimensions is straightforward. We postulate that motion can always be decomposed into fast and slow modes. The slow modes, positions of cluster centers, are adiabatic invariants, which settle down to maximize separation between them just out of Machian reach and retain fractonic behavior. The fast modes representing relative motion within each cluster lose their fractonic character and behave as regular interacting particles within a strong confining external potential generated by the cluster centers and their divergent momenta. For higher dimensions where momenta have a larger space to branch out, this naturally leads to a nearly regular, close-packing arrangement with small deviations from regularity given by $\xi$. Clustering also results in alignment of the direction of momenta within each cluster and is visualized by attaching an arrow corresponding to the direction of the momentum to each particle in \cref{fig:main}(e).

    The various clustering choices are attractors~\cite{AP2023NRFractons} in the position-velocity space of solutions. To see this, notice that from the above calculation for three particles, keeping the essential dynamics for the fast coordinates $x,p$ fixed i.e. leading to the same amplitude $\xi$, we see that various initial configurations for the slow variables $X(0), P(0)$ all lead, at late times to $\dot{X} \rightarrow 0$ and $3X = \pm \left( 1+ \xi \right)$. This generalizes to arbitrary numbers of particles. The space of the attractors, which is an unbounded continuous space (see \cref{app:unboundedness}) can be classified by the locations and membership of the clusters.

    \section{Failure and success of statistical mechanics} We can study the structure of phase space explored by the fracton system and how the breaking of ergodicity occurs from the point of view of statistical mechanics. Let us begin by writing down the partition function in the canonical prescription for the one-dimensional Hamiltonian in \cref{eq:H_dipole} with conservation laws imposed,
    \begin{multline}
        \cZ =\int \prod_{j=1}^N dx_j \delta(\sum_j x_j - X_{\rm tot} ) \\ \int \prod_{j=1}^N dp_j   \delta(\sum_j p_j-P_{\rm tot}) e^{-\beta H }.
    \end{multline}
    Since $H$ is conveniently quadratic in momenta, we can consider integrating them out to generate a statistical probability for the positions of the particles
    \begin{multline}
        P(x_1,\ldots,x_N) = \int \prod_{j=1}^N dp_j  \delta (\sum_j p_j-P_{\rm tot}) e^{-\beta H } \\= \sqrt{ \frac{(2\pi)^N}{\beta \det'L}}. \label{eq:P(x)}
    \end{multline}
    where, we have expressed the Hamiltonian as
    \begin{equation}
        H = \sum_{j<k} \frac{(p_j - p_k)^2}{2} K(x_j - x_k) = \frac{1}{2} \sum_{j,k} p_j L_{jk} p_k. \label{eq:H_graph}
    \end{equation}
    The probability distribution depends on the nature of the eigenvalues of $L$ that assumes a nice form if we consider the limiting form of the mobility function shown in \cref{eq:K_box}. Now, the system can be given the interpretation of an undirected simple graph $G$, where the particles $1\ldots N$ label the vertices of the graph, $V(G)$ and the edges $E(G)$ correspond to pairs $(i,j)$ such that $K(x_i -x_j) = 1$. The matrix $L$ in \cref{eq:H_graph} is the Laplacian of $G$~\cite{Chung1997spectralGraphTheory},
    \begin{equation}
        L(G) = D(G) - A(G). \label{eq:Laplacian}
    \end{equation}
    where, $D(G)$ is the degree matrix of the graph with only diagonal elements $D_{ii}$ containing the degree of the vertex $i$ and $A(G)$ is the adjacency matrix of the graph. The disconnected components of the graph $G$ correspond to clusters.  A well-known result~\cite{Chung1997spectralGraphTheory} states that the number of connected components of $G$ equals the dimensionality of the nullspace, that is, the number of zero eigenvalues of $L$.  Note that $L$ always has at least one zero eigenvalue even when $G$ has a single component, which is eliminated by $\delta(\sum_j p_j - P_{\rm tot})$ which imposes momentum conservation. The expression $\det' L$ in \cref{eq:P(x)} denotes the product of all other eigenvalues that may or may not be zero. For a connected graph, that is, when all particles that form a single cluster, $P(x_1,\ldots,x_N)$ is finite. \emph{Any} other configuration that leads to a graph with multipole components results in a Laplacian $L$ with zero eigenvalues and thus a divergent \cref{eq:P(x)}. Hence, statistical mechanics fails which is consistent, morally, with its breakdown viewed from dynamics.
    
\begin{figure}
            \centering
            \includegraphics[width=8.6cm]{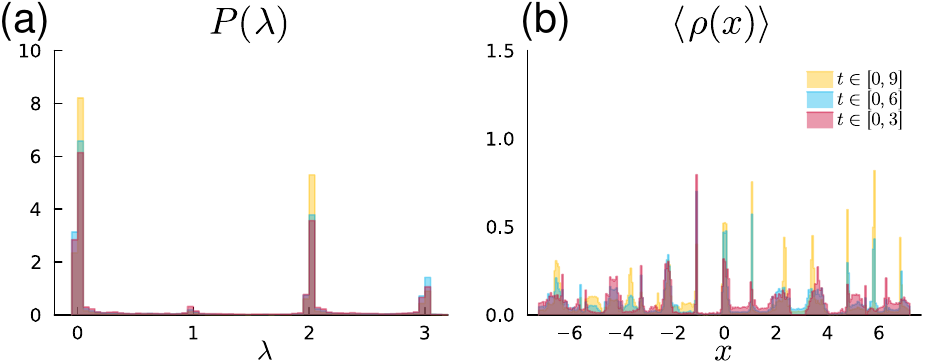}
            \caption{
            (a): Histogram of the eigenvalues of the  $L$ matrix defined in \cref{eq:H_graph} for the same simulation as \cref{fig:main}(a). A large peak at $\lambda=0$ is observed for late times corresponding to the formation of clusters. 
            (b): Particle densities binned over the same time windows. At late times, the density is peaked in each cluster, indicating translation symmetry breaking.
            }
            \label{fig:eigenvalues_density}
        \end{figure}

    However, not all infinities are the same and we can extract useful guidance from statistical mechanics by splitting the $N-1$ dimensional space of positions $x_1,\ldots,x_N$ with a fixed center of mass into different sectors depending on the connectivity of the graph and postulating that sectors with more zero modes will appear more frequently in time as the system evolves. This works surprisingly well, as shown in \cref{fig:eigenvalues_density}(a) for a typical trajectory. However, one cannot use this reasoning to confidently predict the final state---else the high density big-bang state would necessarily lead to a system spanning crystal and we have already noted that it does not. The reasoning also generalizes to higher dimensions, where, although the phase-space variables are vector-valued, the graph-theoretic interpretation is the same.

    At this point, the reader may already have recalled the phenomenon of order-by-disorder (OBD) (see \cite{Villain1980OBD,Shender1982_OBD_antiferromagnetic,ChalkerHoldsworthShender_OBD_PhysRevLett.68.855} and especially \cite{MoessnerChalkerOBDPhysRevLett.80.2929,ChalkerOBD2011}) where geometric frustration leads to a large manifold of ground states, but entropy coming from integration in orthogonal directions selects configurations that host the maximum number of soft modes leading to unexpected order. While the family resemblance to our rationalization of the selection of attractors is very compelling, it is important to note that OBD is invoked in ergodic systems and cannot lead to a violation of the HMWC theorem. In more detail, while the dynamics absolutely takes advantage of the unbounded energy hypersurfaces in phase space there is no sense in which it is ergodic on them that would justify the entropy counting within the traditional ergodic framework.
    
    \section{Higher multipole conservation}
    \begin{figure}[!ht]
	\includegraphics[width=8.6cm]{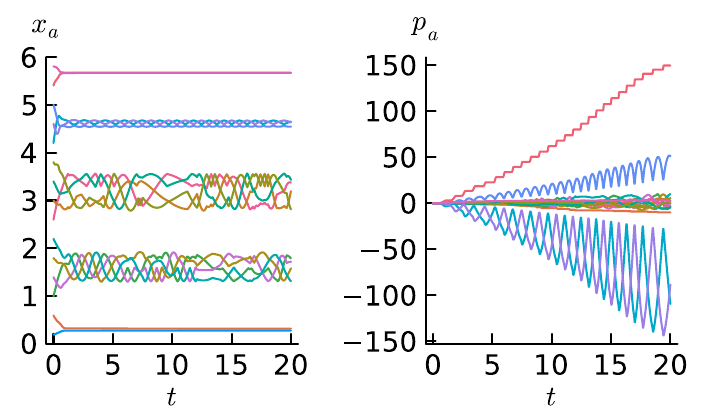}
	\caption{Quadrupole $Q_2$ conserving system of 15 particles ($\rho = 2.5$), starting from uniform density.}
    \label{fig:quadrupole}
\end{figure}
   We now generalize the Hamiltonian in \cref{eq:H_dipole} which is invariant under translations and dipole symmetry to those with translations and multipole symmetry. We keep to one dimension for simplicity, where the conserved multipole moment is
    \begin{equation}
        Q_{\ell} \equiv \sum_j x_j^\ell. \label{eq:Ql}
    \end{equation}
    To begin, let us note that the Poisson bracket between $Q_\ell$ and the total momentum $P = \sum_j p_j$ is non-vanishing and symmetry generators satisfy the classical multipole algebra~\cite{Gromov_Multipole_PhysRevX.9.031035},
    \begin{equation}
        \{Q_\ell,P\}  = \ell Q_{\ell-1},~\{Q_{\ell}, Q_{\ell'} \} =0. 
    \end{equation}
    From Jacobi's identity, we see that conservation of $Q_{\ell}$ and $P$ imposes conservation of all $Q_{1},\ldots,Q_\ell$.
    \begin{multline}
        \{\{H,Q_{\ell}\},P\}+\{\{P,H \},Q_{\ell} \} +\{\{Q_l,P\},H \} =0 \\\implies \{H,Q_{\ell-1}\} = 0.
    \end{multline}
    A Hamiltonian with all these symmetries is
    \begin{equation}
        H = \sum_{a_1,\ldots,a_{\ell+1}=1}^N L^2(a_1,\ldots,a_{\ell+1}) K(x_{a_1},\ldots,x_{a_{\ell+1}}) \label{eq:General Hamiltonian}
    \end{equation}
    where $L(a_0,\ldots,a_{ \ell})$ is an $\ell+1$ body term defined as 
    \begin{equation}
        L(a_0,\ldots,a_{ \ell }) = \sum_{\alpha_k\in a_0,\ldots,a_{ \ell} } \epsilon_{\alpha_0\ldots\alpha_{\ell}}~ p_{\alpha_0} \prod_{k = 1}^\ell x_{\alpha_k}^{\ell - k}. 
    \end{equation}
    $\epsilon_{\alpha_0,\ldots,\alpha_\ell}$ is the Levi-Civita tensor whose elements are determined through its total antisymmetry property via a choice for one of the elements, say $\epsilon_{a_0 \ldots a_{\ell}} = +1$. This choice does not matter because $L$ is squared in \cref{eq:General Hamiltonian}. $K(x_0,\ldots,x_\ell)$ is any translationally invariant term that imposes locality on the $\ell+1$ body term. A suitable form for $K$ is  
    \begin{equation}
        K(x_1,\ldots,x_k) = \prod_{a<b = 1}^k K(x_a - x_b).
    \end{equation}
    It is easy to verify that \cref{eq:General Hamiltonian} is invariant under translations and the symmetries generated by $\{Q_1,\ldots Q_\ell\}$ i.e. 
    \begin{align}
    x_a &\mapsto x_a + \alpha,\\
    p_a &\mapsto p_a + \sum_{k=0}^{\ell - 1} \beta_k x_a^{k}.    
    \end{align}
    For concreteness, let us write down the form of $L$ corresponding to \cref{eq:General Hamiltonian} for $\ell = 2$ 
    \begin{multline}
         L(a,b,c) =p_a (x_b- x_c)  + p_b (x_c - x_a) + p_c (x_a - x_b). \nonumber
    \end{multline}
    As shown in \cref{fig:quadrupole}, clustering and ergodicity breaking is observed in the dynamics of quadrupole-conserving fractons qualitatively similar to dipole conserving ones.
    
    \Cref{eq:General Hamiltonian} generates a more complex flavour of Machian dynamics. While motion of \cref{eq:H_dipole} requires the presence of at least two proximate particles, \cref{eq:General Hamiltonian} requires at least $\ell+1$ particles. We may ask whether an alternative Hamiltonian with fewer interacting particles may be found with the same symmetries. We now present a geometric argument to show that this is not so. The important observation is that when a Hamiltonian is built of local terms, each term should be independently symmetric, which places constraints on the available space to explore. Let us begin with dipole conservation and consider a k-body term that preserves it. That it, any dynamics induced by the local term preserves
\begin{equation}
   \sum_{a=1}^k x_a  = Q_1 \label{eq:k-dipole}
\end{equation}
This tells us that dynamics occurs along a $k-1$ dimensional hypersurface in $\bR^k$ preserving \cref{eq:k-dipole}. It is only for $k\ge 2$ that the dynamics can be non-trivial. For $k=1$ for instance, we get a constraint $x_1 = Q_1$ with no additional freedom and thus no dynamics. Thus, we recover the fact that dipole-conserving Hamiltonians are built out of at least two-body terms. For systems that conserve dipole and quadrupole moments, a local k-body term needs to satisfy \cref{eq:k-dipole} as well as an additional constraint,
\begin{equation}
    \sum_{a=1}^k x_a^2 = Q_2. \label{eq:k-quadrupole} 
\end{equation}
We see that for \cref{eq:k-dipole,eq:k-quadrupole} describe a $k-2$ dimensional hypersurface in $\bR^k$ where dynamics can occur. However, for $k=2$, when a solution exists, it yields a discrete set of points with no freedom for dynamics. Thus, a local term generating non-trivial dynamics occurs for $k=3$ i.e. a 3-body term. This generalizes to general $\ell$. The local term needs to conserve $\ell$ conservation laws $Q_1,\ldots,Q_{\ell}$ with
\begin{equation}
    \sum_{a=1}^k x_a^m = Q_m. \label{eq:k-multipole}
\end{equation}
The only way for the system to have dynamics is if the set of equations in \cref{eq:k-multipole} are overdetermined i.e. the number of variables are $\ge \ell+1$. The motion then occurs on the $k-\ell$ dimensional hypersurface formed by the intersection of \cref{eq:k-multipole}.

   \section{In closing}    We have presented a novel, robust setting for ergodicity breaking in classical systems where symmetries and locality lead to dynamical non-equilibrium steady states governed by attractors in position-velocity space that evade both Liouville and Hohenberg-Mermin-Wagner-Coleman theorems. 
   
   At the classical level, the next obvious task is to gauge the system and study the resulting dynamics. This will require us to work in two- and higher-dimensions. In our previous work~\cite{AP2023NRFractons}, we showed that strong attractions can qualitatively change the two-particle dynamics. We have not found any similar change for the many-particle problem for generic initial conditions, as the effect of momentum divergence dominates the effect of any interaction at late times, leading to robust clustering properties.   
   Nevertheless, we cannot rule out the existence of interactions that break this picture for fine-tuned, e.g. crystalline, initial conditions, and it would be useful to clarify this. Quantizing our system and then comparing what we find with available results on lattice quantum systems is another natural task. We noted that ergodicity breaking at high particle densities is not observed in quantum lattice systems~\cite{RahulNandkishore_FractonLocalizationPhysRevX.9.021003,SalaRakovskyVerresenKnapPollmann_FragmentationPhysRevX.10.011047,KhemaniHermeleNandkishore_Shattering_PhysRevB.101.174204,MorningstayKhemaniHust_Thermalization_PhysRevB.101.214205,SkinnerPozderac_Thermalization_2023}.  On a related note, the reader might wonder what the effect is of relaxing the strict compact nature of the mobility function in \cref{eq:K}. A form with exponential tails was studied in \cite{AP2023NRFractons} and was shown to produce clustering which we have numerically checked persists for a large number of particles as well. This is again in contrast to quantum lattice models, where an equivalent modification is expected to restore ergodicity. It would be interesting to further explore the qualitative changes in physics when passing from a lattice to continuum, consistent with the phenomenon of UV-IR mixing ~\cite{GorantlaLamSeiberg_UVIR_PhysRevB.104.235116,YouMoessner_UVIR_PhysRevB.106.115145,Minwalla_UVIR_2000} known to occur in fracton systems.  Finally, it would be useful to make connections to realistic systems where our results can potentially be observed. A promising setting is the presence of strong tilted fields ~\cite{Tilted2021scherg2021observing,Tilted2021morong2021observation,StarkMBL2021PhysRevLett.127.240502} and harmonic traps~\cite{BagchiHardrods2023unusual} that may dynamically produce the conservation of multipole moments.

        \medskip 
    \noindent\emph{Acknowledgments}: We thank John Chalker, Siddharth Parameswaran, David Logan, Sanjay Moudgalya, Michael Knap, Frank Pollmann, Jonathan Classen-Howes,  Riccardo Sense, Rahul Nandkishore for helpful discussions and Alain Goriely for collaboration on related work~\cite{AP2023NRFractons}.  A.P. was supported by the European Research Council under the European Union Horizon 2020 Research and Innovation Programme, Grant Agreement No. 804213-TMCS and the Engineering and Physical Sciences Research Council, Grant number EP/S020527/1. S.L.S. and Y.S. were supported by a Leverhulme Trust International Professorship, Grant Number LIP-202-014. For the purpose of Open Access, the authors have applied a CC BY public copyright license to any Author Accepted Manuscript version arising from this submission. 

\appendix

	\section{Details of the adiabatic approximation}
 \label{app:Adiabatic}
 We give additional details for the adiabatic calculation for the three-particle Machian schismogenesis presented in the main text.
	\subsection{Preliminaries: two fractons on a line}
 \label{app:2_fracton}
	\begin{figure}[!ht]
	\includegraphics[width=8.6cm]{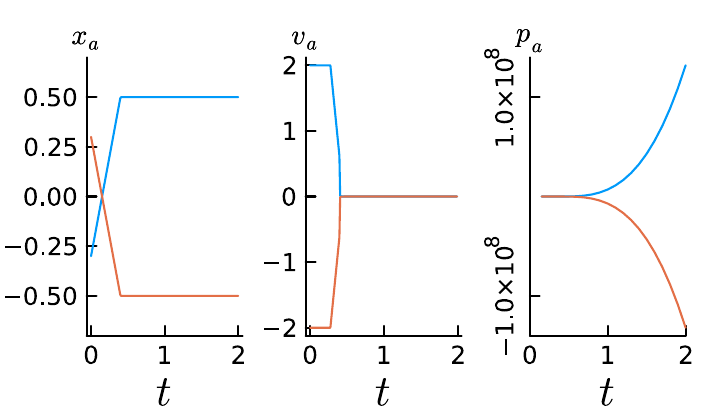}
	\caption{Position, velocity and momentum trajectories for two particles without interactions generated by \cref{eq:Heff2} with $g = 0.01$, $\vec{x}(0) = (-0.3,0.3)$ and $\vec{p}(0) = (1,-1)$.}

	\label{fig:2particle}
	\end{figure}
 
	We begin by with the Hamiltonian for two dipole-conserving fractons on a line 
	\begin{equation}
		H = \frac{(p_1-p_2)^2}{2} K(x_1-x_2) \label{eq:H_2particle}
	\end{equation}
	Using the canonical transformation~\cite{AP2023NRFractons}
        \begin{align}
            X &= \frac{x_1+x_2}{\sqrt{2}}, P = \frac{p_1+p_2}{\sqrt{2}}, \nonumber\\
             x &= \frac{x_1-x_2}{\sqrt{2}}, p = \frac{p_1-p_2}{\sqrt{2}}, \label{eq:canonical_2particle}
        \end{align}
  we can eliminate $X,P$ and write \cref{eq:H_2particle} as
	\begin{equation}
		H = p^2 K(\sqrt{2} x). \label{eq:Heff2}
	\end{equation}
 	and the equations of motion are
	\begin{equation}
		\dot{x} = 2p K(\sqrt{2} x),~ \dot{p} = \sqrt{2} p^2 K'(\sqrt{2} x). \label{eq:eom_2}
	\end{equation}
        \Cref{eq:Heff2} corresponds to a single degree of freedom and its equations of motion in \cref{eq:eom_2} can be solved by quadrature~\cite{AP2023NRFractons}. We consider the form for $K(x)$ is the same considered in the main text i.e.
        \begin{equation}
     K(x) = \begin{cases}
      1 & \mbox{ $ x^2 \leq 1-2g $}\\
      1 - \frac{1}{2g^3}(x^2 -1 +2g)^3 & \mbox{ $ 1-2g < x^2 \leq 1-g $}\\
      - \frac{1}{2g^3}(x^2 -1)^3 & \mbox{ $ 1-g < x^2 < 1 $}\\
      0 & \mbox{ $ x^2 \geq 1 $}.
      \end{cases} \label{eq:K_appendix}
      \end{equation}
    As discussed in \cite{AP2023NRFractons}, under the dynamics generated by \cref{eq:Heff2}, as $t \rightarrow \infty$, $x \rightarrow \pm 1, \dot{x} \rightarrow 0$ and $p \rightarrow \infty$. In other words, the particles separate and become immobile as the momentum difference diverges. The two are related, i.e. freezeout happens as the momenta split up into two branches and grow in magnitude. This is shown in \cref{fig:2particle}.
	\begin{figure}[!ht]
	\includegraphics[width=8.6cm]{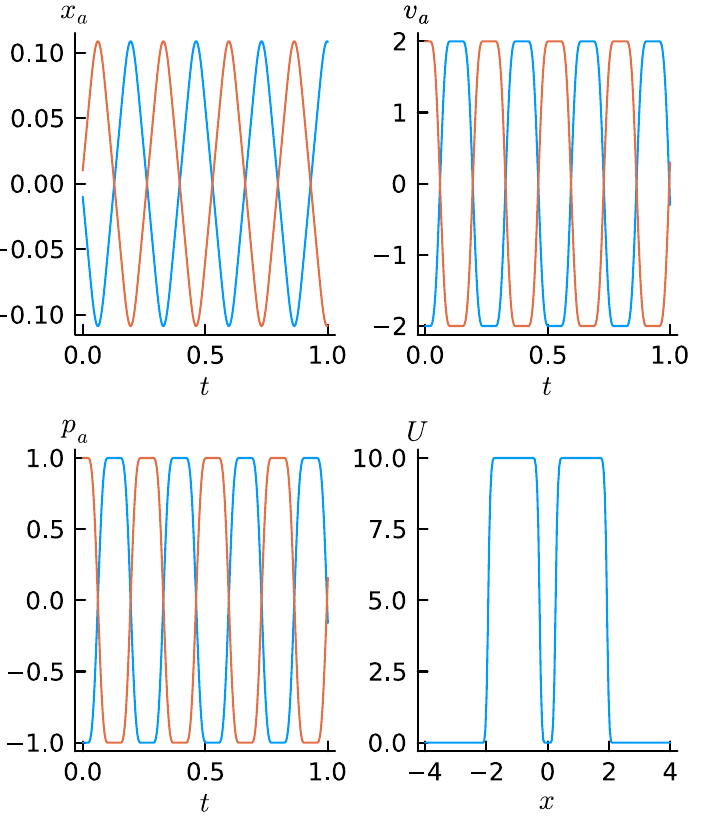}
	\caption{Trajectories for two particles without interactions generated by \cref{eq:Heff2} with $\vec{x}(0) = (-0.01,0.01)$ and $\vec{p}(0) = (-1,1)$. Plot of $U(x)$ used with $\Gamma = 10$, $\xi = 0.1$ and $g=0.3$.}
	\label{fig:2particle_interaction}
\end{figure}	
Let us now add an interaction $U(x_1 - x_2)$ which produces short range attraction 
	\begin{equation}
		U_\xi(x) = \Gamma \left( K(1+\xi+x) + K(1+\xi-x) \right) \label{eq:U}
	\end{equation}
as shown in \cref{fig:2particle_interaction}(d). This particular in \cref{eq:U} will be relevant to us soon. The new Hamiltonian is 
	\begin{equation}
		H =  p^2 K(\sqrt{2} x) + U_\xi(\sqrt{2} x) \label{eq:H_2_int}
	\end{equation} 
	and the equations of motion are
	\begin{equation}
		\dot{x} = 2p K(\sqrt{2} x),~ \dot{p} = \sqrt{2} p^2 K'(\sqrt{2} x) +\sqrt{2} U'_\xi(\sqrt{2}x). \label{eq:eom_2_interaction}
	\end{equation}
	The two-particle interaction in \cref{eq:H_2_int} corresponds to a background potential for the effective single degree of freedom.  For large $|U|$ , an initially confined particle remains so and does not explore large values of $x$. Thus $K \approx 1$ and $K' \approx 0$ giving us the effective equations
	\begin{equation}
		\dot{x} \approx 2p,~ \dot{p} \approx \sqrt{2} U'_\xi(\sqrt{2}x). 
	\end{equation}
	In other words, the system of $(x,p)$ behaves like an ordinary particle of mass $m =0.5$, governed by the effective Hamiltonian
	\begin{equation}
		H_{\rm eff} \approx p^2 + U_\xi(\sqrt{2}x). \label{eq:2particle_ordinary}
	\end{equation}
	For initial conditions starting near $x=0$, the system oscillates about $x=0$ as shown in \cref{fig:3particle}. 
	\subsection{3 particles on a line}
		\begin{figure}[!ht]
		\includegraphics[width=8.6cm]{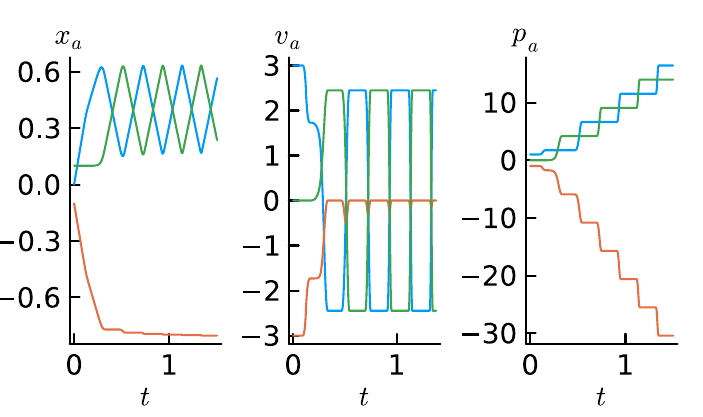}
		\caption{Trajectories for three particles without interactions generated by \cref{eq:3particleH} with $g=0.3$, $\vec{x}(0) = (0,-0.1,0.1)$ and $\vec{p}(0) = (1,-1,0)$. }
		\label{fig:3particle}
	\end{figure}
	Let us now consider three particles. The Hamiltonian we are interested in is
	\begin{multline}
		H = \frac{(p_1 - p_2)^2}{2} K(x_1 - x_2) +\frac{(p_1 - p_3)^2}{2} K(x_1 - x_3) \\+\frac{(p_2 - p_3)^2}{2} K(x_2 - x_3)  \label{eq:3particleH}
	\end{multline}
and the equations of motion are
\begin{align}
	\dot{x}_1 &=+ (p_1 - p_2) K(x_1 - x_2) +(p_1 - p_3) K(x_1 - x_3), \nonumber \\
	\dot{x}_2 &= -(p_1 - p_2) K(x_1 - x_2) +(p_2 - p_3) K(x_2 - x_3),\nonumber \\
	\dot{x}_3 &= -(p_1 - p_3) K(x_1 - x_3) -(p_2 - p_3) K(x_2 - x_3),\nonumber \\	
	\dot{p}_1 &= +\frac{(p_1 - p_2)^2}{2} K'(x_1 - x_2) + \frac{(p_1 - p_2)^3}{2} K'(x_1 - x_3)\nonumber,\\
	\dot{p}_2 &= -\frac{(p_1 - p_2)^2}{2} K'(x_1 - x_2) + \frac{(p_2 - p_3)^3}{2} K'(x_2 - x_3),\nonumber	\\
	\dot{p}_3 &= -\frac{(p_1 - p_3)^2}{2} K'(x_1 - x_3) - \frac{(p_2 - p_3)^3}{2} K'(x_2 - x_3). \label{eq:3_eom}		
\end{align}
	Generically, the dynamics for this Hamiltonian looks as shown in \cref{fig:3particle}. We see that after a brief, transient period, two particles (1,2) form a cluster and settle into indefinite oscillations, while the third becomes motionless. An exact solution for this dynamics was presented in \cite{AP2023NRFractons} for the limiting form $g \rightarrow 0$ in \cref{eq:K_appendix}.  We want to understand this behavior qualitatively and determine the mechanism of clustering. The key is in the nature of momentum-space dynamics. The clustering in position space occurs just as the three momenta branch out into two clusters. 
	
	We see that there are two time scales of motion. The first fast scale is the dynamics of particles 1,2 in the first cluster relative to their center of mass and the second, slow scale is in the motion of the two centers of mass. Now we decompose the coordinates to incorporate this separation of scale. Before we do this, notice that the coordinates are not independent and are constrained by symmetries:
	\begin{equation}
		\dot{x}_1 + \dot{x}_2 + \dot{x}_3 = 		\dot{p}_1 + \dot{p}_2 + \dot{p}_3 =0.
	\end{equation}
	We can use this to eliminate one pair of canonically conjugate phase space variables. Without loss of generality, we set $x_1 + x_2 + x_3 = p_1 + p_2 + p_3 = 0$ and define the following coordinates
	\begin{align}
		x_1 = X + x,~ x_2 = X-x,~ x_3 = -2X,\\
		p_1 = P + p,~ p_2 = P-p,~ p_3 = -2P.		
	\end{align}
	Using this, the equations of motion shown in \cref{eq:3_eom} can be rewritten as
 \begin{widetext}
	\begin{align}
		2\dot{X} &= \dot{x}_1 + \dot{x}_2 = (3P +p) K(3X + x) + (3P-p) K(3X-x),\\
		2\dot{x} &= \dot{x}_1 - \dot{x}_2  = (4p) K(2x) + (3P +p) K(3X + x) - (3P-p) K(3X-x),\\
		2\dot{P} &= \dot{p}_1 + \dot{p}_2 = -\frac{(3P +p)^2}{2} K'(3X + x) - \frac{(3P-p)^2}{2} K'(3X-x), \\
		2\dot{p} &= \dot{p}_1 - \dot{p}_2  = -(2p)^2 K'(2x) - \frac{(3P +p)^2}{2} K'(3X + x) + \frac{(3P-p)^2}{2} K'(3X-x).		
	\end{align}     
 \end{widetext}

 The separation of scales allows us to employ an adiabatic approximation~\cite{LandauVol11982mechanics}. We solve the equations of motion for $x,p$ assuming $X,P$ are constant and then feed back a time-averaged solution to solve for $X,P$. Let us begin with the latter, by assuming that
 \begin{equation}
  P>>p, ~x<1,~2>3X>1. \label{eq:adiabatic_assumptions}   
 \end{equation}
This will be self-consistently justified later. The equations of motion for $x$ and $p$ simplify to
 \begin{align}
    \dot{x} &\approx  2p K(2x), \nonumber \\
    \dot{p} &\approx -2 p^2 K'(2x) - \frac{9P^2}{4}\left( K'(3X + x) -  K'(3X-x)\right).		\label{eq:3_adiabatic_xp_1}
 \end{align}
 \Cref{eq:3_adiabatic_xp_1} is generated by an effective Hamiltonian
 \begin{equation}
     H_{\rm eff} \approx p^2 K(2x) + U_{X,P}(x). \label{eq:Heff_xp_1}
 \end{equation}
 where 
 \begin{equation}
     U_{X,P}(x) = \frac{9 P^2}{4} \left(K(3X+x) + K(3X-x) \right).
 \end{equation}
 We see that \cref{eq:Heff_xp_1} is qualitatively the same as \cref{eq:H_2_int} if $3X \rightarrow 1+\xi$. Thus, $x$ exhibits rapid motion with an amplitude $\sim \xi$ which can be described by simplifying \cref{eq:Heff_xp_1} further by setting $ K(2x) \approx 1$ giving us an effective Hamiltonian of an ordinary particle in a confining potential, the same as in \cref{eq:2particle_ordinary}
  \begin{equation}
     H_{\rm eff} \approx p^2 + U_{X,P}(x). \label{eq:Heff_xp}
 \end{equation}

 \begin{figure}[!ht]
 \includegraphics[width=8.6cm]{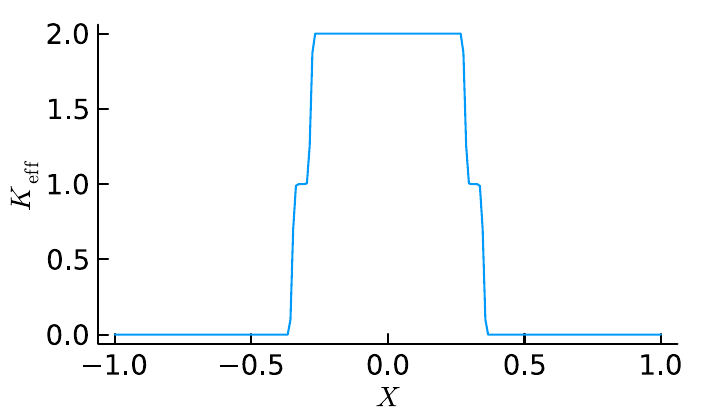}
	\caption{$K_{\rm eff}$ defined in  \cref{eq:Keff_supp} with $\xi = 0.1$ and $g=0.1$. \label{fig:Keff}} 
\end{figure}

 Let us now consider the dynamics of the slow degrees of freedom $X,P$. Under the assumptions in \cref{eq:adiabatic_assumptions} and incorporating the fast solution by simply replacing $x$ by its amplitude $\xi$ we get
	\begin{align}
		\dot{X} &\approx \frac{3P}{2} \left(K(3X + \xi) +  K(3X-\xi) \right), \nonumber\\
		\dot{P} &\approx  -\frac{3P^2}{4}\frac{\partial \left( K(3X + \xi) +  K(3X-\xi)\right)}{\partial X}.\label{eq:XP}
	\end{align}
 
	In other words, the dynamics for $X,P$ are generated by the effective Hamiltonian 
	\begin{equation}
		H_{\rm eff} \approx \frac{3P^2}{4} K_{\rm eff}(X). \label{eq:Heff_XP}
	\end{equation}
	where, 
	\begin{equation}
		K_{\rm eff}(X) \equiv K(3X + \xi) +  K(3X-\xi). \label{eq:Keff_supp}
	\end{equation}
	$K_{\rm eff}$ is shown in \cref{fig:Keff}.
	Remarkably, \cref{eq:Heff_XP} is qualitatively the same as \cref{eq:Heff2}. From the discussion in \cref{app:2_fracton}, we know that under dynamics at late times, $X$ settles down at the `edges' where $K_{\rm eff}$ vanishes i.e. $3X = \pm (1+\xi) >1$ and $P \rightarrow \infty$. This self-consistently justifies the assumptions made in \cref{eq:adiabatic_assumptions} and the adiabatic approximation should reproduce late-time dynamics reliably.

\section{Unboundedness of the space of attractors}
\label{app:unboundedness}

\begin{figure}[!ht]
    \centering
    \includegraphics[width=8.6cm]{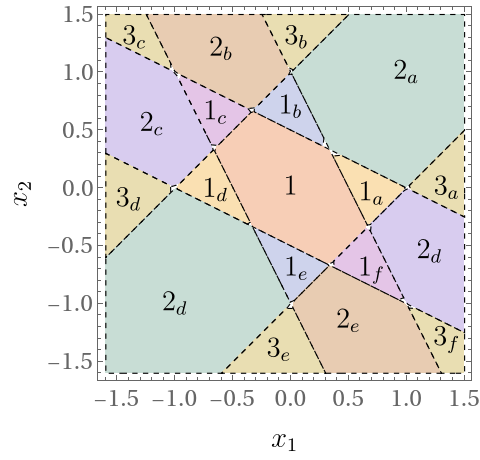}
    \caption{The position space divided into different regions marked by the number of clusters. We see that $2_a - 2_f$ and $3_a - 3_f$ which have two and three clusters respectively are unbounded.}
    \label{fig:3_Cluster}
\end{figure}
In the main text, we stated that the space of attractors i.e. clustering configurations  is unbounded leading to a divergent statistical probability whereas the space of configurations forming a single cluster with a finite statistical probability is finite. Let us visualize this for the space of three particles. Imposing the conservation law $x_1 + x_2 + x_3 = 0$, the resulting two-dimensional position space is divided as shown in \cref{fig:3_Cluster}. The region marked $1$ represents the region where all particles are within Machian reach, that is, $|x_a - x_b| \le 1$ for all $a,b \in 1,2,3$. The region marked $1_a - 1_f$ represents the region where all three particles form a single cluster with pairwise Machian reach, i.e. $|x_1 - x_2| <1, ~|x_2 - x_3| <1,|x_1 - x_3| >1$ and other permutations. The finite region consisting of $1$ and $1_a - 1_f$ corresponds to a single cluster leading to a finite statistical probability $P(x_1,x_2,x_3)$. The regions $2_a - 2_f$ contain two clusters where only one pair is within Machian reach i.e. $|x_1 - x_2| <1, ~|x_2 - x_3| >1,|x_1 - x_3| >1$ and other permutations and the regions $3_a - 3_f$ contain three clusters where all particles are out of Machian reach i.e. $|x_a - x_b| >1$ for all $a,b \in 1,2,3$. We see that the regions $2_a - 2_f$ and $3_a - 3_f$ are unbounded, where $P(x_1,x_2,x_3)$ diverges. Dynamically, we see that when the particle starts in $1$ or $1_a - 1_f$, it is most likely to encounter $2_a - 2_f$ which represents the steady states described in the main text where two particles form a cluster with the third out of reach.

\section{Initial condition dependence}
\label{app:initial_condition_dependence}

In the main text, we presented trajectories in 1 and 2 dimensions for systems starting from configurations of approximately uniform density (and $\rho > 1$). This dynamically leads to the emergent crystallisation, filling the volume.
However this does not occur for some specific initial conditions.
For example, starting a large number of particles in a small volume, such that all $K(x_i - x_j) =1$, does \emph{not} expand to a uniform crystal. 

\begin{figure}[H]
	\includegraphics[width=8.6cm]{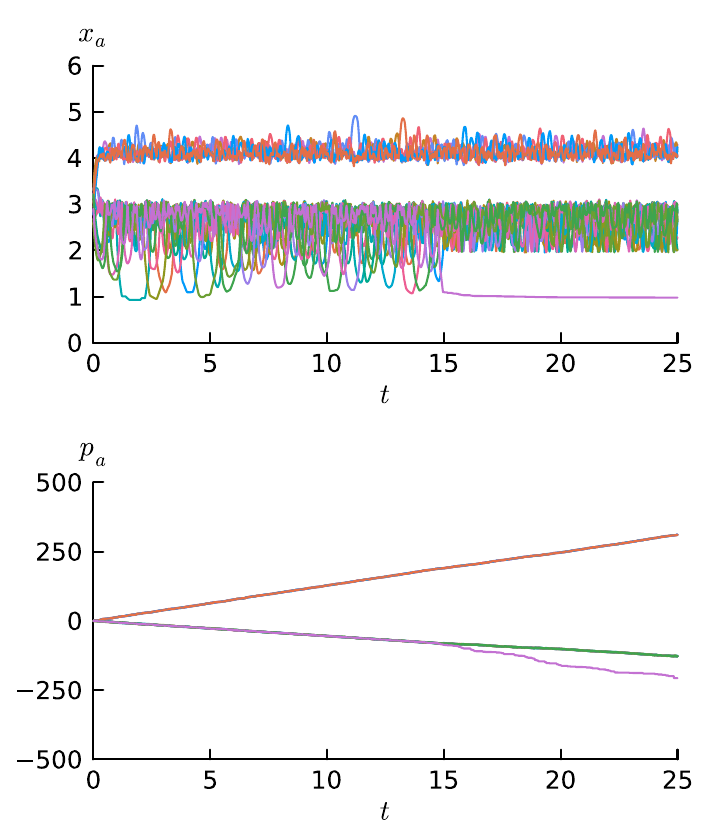}
	\caption{20 particles in 1D starting within close proximity.}
    \label{fig:spread_centre}
\end{figure}

Shown in \cref{fig:spread_centre}, an single cluster evolves into only 3 clusters at late times, with a large number of particles remaining in the central cluster.
This is a general result for such initial conditions: even if the number of particles is taken to be very large, few clusters will form, as large energy barriers form between the individual clusters.

\bibliography{references}

\end{document}